\documentclass[11pt,aps,preprintnumbers,nofootinbib,superscriptaddress,notitlepage,longbibliography]{revtex4-2}

\usepackage[T1]{fontenc}
\usepackage[english]{babel}
\usepackage{stix}
\usepackage{epsfig}
\usepackage{amsmath,bm,bbm}
\usepackage{graphicx}
\usepackage{verbatim}
\usepackage{csquotes}
\usepackage{simplewick}
\usepackage{appendix}
\usepackage[usenames,dvipsnames,svgnames]{xcolor}
\usepackage[normalem]{ulem}
\usepackage{hyperref}
\usepackage{cleveref}

\usepackage[framemethod=default]{mdframed}
\newmdenv[skipabove=7pt,
skipbelow=7pt,
rightline=false,
leftline=false,
topline=false,
bottomline=false,
backgroundcolor=gray!10,
linecolor=gray,
innerleftmargin=5pt,
innerrightmargin=5pt,
innertopmargin=5pt,
innerbottommargin=5pt,
leftmargin=0cm,
rightmargin=0cm,
linewidth=4pt]{eBox}

\newcommand{\md}{\mathrm{d}}

\begin{document}

\title{Connecting Relativistic MOND Theories with Mimetic Gravity }

\author{\textsc{Guillem Dom\`enech }}
    \email{{guillem.domenech}@{itp.uni-hannover.de}}
    \affiliation{ Institute for Theoretical Physics, Leibniz University Hannover, Appelstraße 2, 30167 Hannover, Germany}
    \affiliation{ Max-Planck-Institut für Gravitationsphysik, Albert-Einstein-Institut, 30167 Hannover, Germany}

\author{\textsc{Alexander Ganz}}
    \email{alexander.ganz@itp.uni-hannover.de}
    \affiliation{ Institute for Theoretical Physics, Leibniz University Hannover, Appelstraße 2, 30167 Hannover, Germany}

\begin{abstract}
    We find a connection between relativistic Modified Newtonian Dynamics (MOND) theories and (scalar) mimetic gravity. We first demonstrate that any relativistic MOND model featuring a unit-timelike vector field, such as TeVeS or Aether-scalar-tensor theory, can be embedded within a conformal/disformal-invariant framework. Gauge fixing the conformal/disformal symmetry amounts to imposing a constraint on the norm of the vector, the scalar field or the cross contraction. Notably, we find that these constraints can be interchanged as long as the vector and scalar fields remain timelike. This means that relativistic MOND theories may be recasted as a mimetic gravity theory. 
    Lastly, by constructing the fundamental building blocks of a conformal-invariant scalar-vector-tensor theory, we establish a new framework for developing relativistic MOND theories. This perspective offers deeper insight into how non-invertible disformal transformations and conformal/disformal symmetries serve as fundamental principles in constructing viable alternatives to dark matter.
\end{abstract}

\maketitle

\section{Introduction}

Despite many promising candidates, there is still no evidence for particle dark matter. It is thus plausible that the observed effects associated with dark matter arise instead from modifications of gravity. A typical indication of dark matter comes from observations of galaxy rotation curves, which cannot be explained by visible baryonic matter alone \cite{1980ApJ...238..471R}. Nonetheless, the motion of stars could also be explained by modifying Newton’s law at small accelerations, a framework known as Modified Newtonian Dynamics (MOND); see, e.g., Refs.~\cite{1983ApJ...270..365M,1984ApJ...286....7B}. See also Ref.~\cite{Bertone:2016nfn} for a review of the history of dark matter. However, one needs a relativistic completion of MOND to study its cosmology. Thus, several relativistic theories that reproduce MOND-like behavior in the weak-field regime have been proposed; see, e.g., Ref.~\cite{Famaey:2011kh} for a review. Note that, though, in a bottom-up approach, there could be many various ways of building a relativistic MOND theory.

A well-known example of a relativistic extension of MOND is the {\it Tensor-Vector-Scalar theory} (TeVeS) \cite{Bekenstein:2004ne}. However, the original TeVeS theory predicts a different propagation speed of gravitational waves than that of light, which is in conflict with current constraints from the observation of a neutron star merger with electromagnetic counterpart \cite{LIGOScientific:2017vwq,LIGOScientific:2017zic}. For this reason, Ref.~\cite{Skordis:2020eui} proposed a new model based on a generalized class of TeVeS-like theories \cite{Skordis:2019fxt}, where gravitational waves propagate at the speed of light. This model has been dubbed the {\it Aether-scalar-tensor} theory, or AeST theory for short. Both TeVeS and AeST theories feature a dynamical scalar field and a unit-timelike vector field, the norm of the latter being enforced via a Lagrange multiplier. Such a constraint has been extensively analyzed in the context of Einstein-Aether gravity \cite{Jacobson:2000xp} to explore the consequences of a preferred reference frame. Similar constraint structures have also been widely studied in the construction of perfect fluids \cite{PhysRevD.2.2762,Brown:1992kc,Brown:1994py}.

Interestingly, there is a subset of scalar-tensor theories with a similar constraint structure. In the so-called mimetic gravity, one imposes time-like gradients of a scalar field.
Mimetic gravity is particularly interesting as it reproduces the behavior of pressureless dust \cite{Lim:2010yk,Chamseddine:2013kea}, which could explain the dark matter in the universe  (see also Refs.~\cite{Chamseddine:2014vna,Mirzagholi:2014ifa} and Ref.~\cite{Sebastiani:2016ras} for a review). Curiously, the mimetic constraint also emerges from a non-invertible conformal transformation \cite{Chamseddine:2013kea,Golovnev:2013jxa,Barvinsky:2013mea}, from a conformal-invariant scalar-tensor theory \cite{Hammer:2015pcx}, and in noncommutative geometry \cite{Chamseddine:2014nxa}.

Mimetic gravity from non-invertible conformal transformation can also be extended to non-invertible disformal transformations,\footnote{Disformal transformations were originally proposed by Bekenstein in Ref.~\cite{Bekenstein:1992pj}.} which depend on both the scalar field and its kinetic term \cite{Deruelle:2014zza,Arroja:2015wpa,Jirousek:2022jhh,Jirousek:2022rym,Golovnev:2022jts}. This has led to increased interest in non-invertible disformal transformations, which have been applied to various settings, including vector fields \cite{Barvinsky:2013mea,Chaichian:2014qba,Jirousek:2018ago,Benisty:2021cin}, p-forms \cite{Gorji:2018okn}, and multi-scalar field models \cite{Firouzjahi:2018xob,Zheng:2022vwm}. Similarly, the conformal invariant formulation of mimetic gravity can equivalently be deformed into disformal symmetry \cite{Jirousek:2022rym,Domenech:2023ryc}. In this case, gauge fixing the conformal or disformal symmetry can be achieved by introducing the mimetic constraint via a Lagrange multiplier term in the action. Unfortunately, general Mimetic gravity suffers from either ghost or gradient instabilities  \cite{Langlois:2018jdg,Takahashi:2017pje,Zheng:2017qfs,Firouzjahi:2017txv,Ijjas:2016pad,Hirano:2017zox,Gorji:2017cai,Ganz:2018mqi,Ganz:2019vre}.

As in mimetic gravity, the vector field in TeVeS is enforced to be time-like through a Lagrange multiplier term. But we may also understand the constraint as a gauge-fixing of conformal symmetry. It then follows that, in such conformal-invariant formulation, the choice of the gauge-fixing condition becomes arbitrary. And since TeVeS and AeST involve a scalar and a vector field, one alternative gauge-fixing choice is to impose the (scalar) mimetic constraint. We may also express TeVes and AeST in terms of a non-invertible conformal (disformal) transformation, as in mimetic gravity. Thus, the two most widely studied modified gravity theories aimed at explaining dark matter--TeVeS and its generalizations, and mimetic gravity--are apparently deeply interconnected. Both share a common conformal or disformal symmetry structure, and both can, therefore, be obtained through non-invertible disformal transformations.

The aim of this paper is to show that the TeVeS constraint can be interchanged by the mimetic constraint and to use that to explore different formulations and generalizations of TeVeS and AeST. We hope that our new formulation of relativistic MOND theories will be helpful in interpreting in a general setting the MOND-like behavior in the weak field limit in terms of known effects in general scalar-tensor theories and generalized Proca theories \cite{Heisenberg:2014rta,Tasinato:2014eka,Tasinato:2014mia,Tasinato:2013oja,Hull:2015uwa,Allys:2015sht,DeFelice:2016yws}, such as additional fifth forces and screenings \cite{Ishak:2018his,Brax:2021wcv,Vardanyan:2023jkm}, like the Vainshtein mechanism \cite{Babichev:2013usa,Kobayashi:2019hrl}. It may also be helpful in finding a frame with simpler equations of motion. We leave this investigation for future work.

This paper is structured as follows. In Sec. \ref{sec:Preliminaries}, we show as an example how the AeST theory can be related to mimetic gravity via a conformal transformation. In Sec. \ref{sec:Conformal_Invariance}, we construct, from first principles, the building blocks for a general conformal-invariant scalar-vector-tensor theory and show how relativistic MOND models emerge as specific cases of this broader framework. In Sec. \ref{sec:Singular_Disformal_transformation}, we explore how conformal and disformal invariant theories can be derived from non-invertible transformations. In Sec. \ref{sec:Aether-Scalar-Gauge-Constraints}, we explore the AeST theory in an FLRW background in the Hamiltonian framework. Finally, we discuss our conclusions in Sec. \ref{sec:Conclusion}. In Appendix \ref{app:Disformal_Invarianz}, we look at the disformal-invariant formulation and demonstrate its equivalence to the conformal-invariant case. It should be noted that we will always denote the constraint on the scalar field gradients as mimetic constraint, while we call the constraint on the vector norm as the TeVeS constraint.

\section{Preliminary example}  
\label{sec:Preliminaries}
Before entering into the details of the construction, symmetries and relations between theories, it may be instructive to look at a simple, illustrative example. Thus, let us show how to formulate the AeST theory in terms of the mimetic constraint. To lay out our notation throughout the paper, consider first a vector-scalar-tensor theory with a general Lagrangian of the form  
\begin{equation}  
\mathcal{L} = \mathcal{L}(A^\mu,\phi,g_{\mu\nu}).  
\end{equation}  
For convenience, we define the following scalar quantities
\begin{equation}  
X \equiv -\nabla^\mu \phi \nabla_\mu \phi~,\quad  
{\cal A} \equiv -A^\mu A_\mu~,\quad  
{\cal Q}\equiv -A_\mu\nabla^\mu \phi ~.  
\end{equation}  
Throughout this work, we assume for simplicity that both the derivative of the scalar field and the vector field are time-like, i.e., $X> 0$, ${\cal A} > 0$, and ${\cal Q} > 0$. Generalization to the space-like case is straightforward. 

The action of the AeST theory reads \cite{Skordis:2020eui} 
\begin{align}  \label{eq:aethereqA}
S_{\text{\AE}} = \int d^4x\,\frac{\sqrt{-g}}{16\pi G}  
\bigg\{  
R -\frac{K_B}{2} F^{\mu\nu} F_{\mu\nu} +& 2(2 - K_B) J^\mu \nabla_\mu \phi - (2 - K_B)\mathcal{Y} \nonumber\\&- \mathcal{F} (\mathcal{Y}, \mathcal{Q}) - \lambda({\cal A} - 1)  
\bigg\}  + S_M(g_{\mu\nu},\Psi_M)~,  
\end{align}  
where $K_B$ is a constant, $F_{\mu\nu}=\nabla_\mu A_\nu-\nabla_\nu A_\mu$ and we identified
\begin{equation}  
J^\mu \equiv A^\alpha \nabla_\alpha A^\mu \qquad {\rm and}\qquad 
\mathcal{Y} \equiv {\cal Q}^2-X ~,
\end{equation}  
in the notation of Ref.~\cite{Skordis:2020eui}. The Lagrange multiplier $\lambda$ enforces a unit time-like norm for the vector. Note that we defined ${\cal Q}$ with the opposite sign as Ref.~\cite{Skordis:2020eui}  for convenience.
The matter sector, denoted by \( S_M \), is minimally coupled to \( g_{\mu\nu} \). It is interesting to note that the vector-scalar interaction can also be rewritten as  
\begin{equation}  \label{eq:theAetherquantities}
J^\mu \nabla_\mu \phi = F^{\mu\nu} A_\mu \nabla_\nu \phi + A^\mu\nabla^\nu A_\mu  \nabla_\nu \phi= F^{\mu\nu} A_\mu \nabla_\nu \phi \Big|_{{\cal A}=1}~, 
\end{equation}  
where in the last step we used the TeVeS constraint. In this form, it is easy to check that the vector-scalar interaction is conformal invariant in the action, just like the term $F^{\mu\nu}F_{\mu\nu}$.  

Let us now show that AeST theory is actually related to mimetic gravity via a regular conformal transformation given by
\begin{equation}  \label{eq:conf_transf}
g_{\mu\nu} = \Omega^2~\tilde{g}_{\mu\nu}\,,
\end{equation}  
where, for the moment, we leave $\Omega^2$ unspecified. After the transformation, the resulting action  reads
\begin{align} \label{eq:newactionaether} 
S_{\text{\AE}} =  \int d^4x\, \frac{\sqrt{-\tilde{g}}}{16\pi G}  
\bigg\{  &
\Omega^2 \tilde{R} +  6 \tilde{g}^{\mu\nu} \partial_\mu \Omega \partial_\nu \Omega - \frac{K_B}{2} F^{\mu\nu} F_{\mu\nu} + 2(2 - K_B) F^{\mu\nu} A_\mu \partial_\nu \phi 
 -(2 - K_B) \left(\tilde{\cal Q}^2-\Omega^2 \tilde X\right)  \nonumber\\ &
 - \Omega^4 \mathcal{F} \left( \frac{\left(\tilde{\cal Q}^2-\Omega^2 \tilde X\right)}{\Omega^4},\frac{\tilde{Q}}{\Omega^2} \right)  
- \lambda \Omega^4 \left( \frac{\tilde{\cal A}}{\Omega^2} - 1 \right)  
\bigg\}  
+ S_M (\Omega^2 \tilde g_{\mu\nu},\Psi_M)\,,
\end{align}
where tilde quantities are given by \eqref{eq:theAetherquantities} but contracted with $\tilde g^{\mu\nu}$. Also, all indexes in the action \eqref{eq:newactionaether} are assumed to be contracted with  $\tilde g^{\mu\nu}$.
Note that the matter sector now has a non-minimal coupling via \( \Omega^2 \).

By fixing a particular functional form of $\Omega^2$ we can change the form of the constraint in Eq.~\eqref{eq:newactionaether}. For example, one possibility is to exchange $\tilde {\cal A}$ by $\tilde {X}$ by choosing
\begin{equation}\label{eq:omegaexample1}  
\Omega^2 = \frac{{\cal A}}{{X}}=\frac{\tilde{\cal A}}{\tilde{X}}~,  
\end{equation}  
which is a non-singular transformation as long as ${\cal A}, X\neq 0$. We also assume ${\cal A}, X>0$ to maintain the signature of the metric. Note that the ratio \({\cal A} /{X} \) is conformal invariant. Inserting $\Omega^2$ from \eqref{eq:omegaexample1} in the action \eqref{eq:newactionaether}, we see that the constraint on the vector field is now replaced by the mimetic constraint \( \tilde{X} = 1 \). We are also free to redefine the Lagrange multiplier to \( \tilde\lambda = \Omega^4 \lambda \). More concretely, the new action reads
\begin{align} 
S_{\text{\AE}} =  \int d^4x\, \frac{\sqrt{-\tilde{g}}}{16\pi G}  
\bigg[  
\tilde{\cal A} \tilde{R} &+ \frac{3}{2 \tilde{\cal A}}\tilde\nabla^\mu \tilde{\cal A} \tilde\nabla_\mu \tilde{\cal A} - \frac{K_B}{2} F^{\mu\nu} F_{\mu\nu} + 2(2 - K_B) F^{\mu\nu} A_\mu \partial_\nu \phi  
-(2 - K_B)  \left( \tilde Q^2 - \tilde X \tilde {\cal A} \right) \nonumber\\ &  - \tilde {\cal A}^2 \mathcal{F} \left( \frac{\tilde Q^2 - \tilde X \tilde {\cal A}}{\tilde X^2},\frac{\tilde{Q}}{\tilde A} \right)  
- \tilde\lambda \left( {\tilde{X}} - 1 \right)  
\bigg]  
+ S_M (\tilde {\cal A}\tilde g_{\mu\nu},\Psi_M)\,.
\label{eq:newactionaether2} 
\end{align}  
The exercise above shows that the AeST theory can be equivalently rewritten in terms of the mimetic constraint combined with a non-minimal coupling of matter to the vector field and a generalized Proca action \cite{Heisenberg:2016eld}. Note that we have only partially used the mimetic constraint to simplify the action. In this formulation, the action is conformal invariant apart from the mimetic constraint. This result is not coincidental; rather, as we show in the next section, the AeST theory can be manifestly reformulated as a conformal-invariant theory, where the constraints on the vector and scalar fields correspond to different gauge-fixing conditions. Lastly, it should be noted that instead of Eq.~\eqref{eq:omegaexample1} we could have chosen
\begin{equation}\label{eq:omegaexample2}  
\Omega^2 = \frac{{\cal A}}{{\cal Q}}=\frac{\tilde{\cal A}}{\widetilde{{\cal Q}}}~\,.  
\end{equation}  
In that case, instead of the mimetic constraint, we would have a constraint given by $\tilde \lambda(\tilde Q+1)$. This is also an equivalent formulation of the AeST theory with ${\cal Q}=1$ instead.

\section{Conformal Invariance}
\label{sec:Conformal_Invariance}

Let us show that TeVeS and AeST theories are part of a broader class of conformal invariant theories. Following Ref.~\cite{Domenech:2023ryc}, we construct basic building blocks which are invariant under conformal transformation. 
First of all, note that due to the presence of a scalar and vector field, we can define a conformal invariant metric as
\begin{align}
\label{eq:Metric_Conformal_Invariance}
    {\cal G}_{\mu\nu} = {\cal A} f(\phi,{\cal X},{\cal P}) g_{\mu\nu}~,
\end{align}
where we defined
\begin{align}
    {\cal X} = \frac{X}{{\cal A}}\qquad {\rm and} \qquad {\cal P}=\frac{{\cal Q}}{{\cal A}}~,
\end{align}
which are two conformal invariant combinations.
The expression can also be derived from requiring that the conformal transformation depending on the scalar and vector field $\Omega(\phi, X,{\cal A},{\cal Q})$ becomes non-invertible (see section \ref{sec:Singular_Disformal_transformation} for more details). This is similar to the original construction of mimetic gravity \cite{Chamseddine:2013kea}.
In the following, we will set $f(\phi,{\cal X},{\cal P})=1$ without loss of generality as we can reconstruct all the conformal invariant operators with any choice of $f(\phi,{\cal X},{\cal P})$.

At leading order, we obtain the conformal invariant building blocks given by
\begin{align}
    {\cal V} = \sqrt{-g} {\cal A}^2~, \qquad {\cal N}^\mu_A = {\cal G}^{\mu\alpha} A_\alpha~, \qquad {\cal N}^\mu_\phi =  {\cal G}^{\mu\alpha} \nabla_\alpha \phi~.
\end{align}
Similarly, we find for the first (second) derivative of the vector (scalar) field that the conformal invariant combinations read
\begin{align}\label{eq:BmunuA}
    {\cal B}_{\mu\nu}^A = & \nabla_{(\mu} A_{\nu)} - \frac{1}{{\cal A}} A_{(\mu}\nabla_{\nu)} {\cal A}  + \frac{1}{2 {\cal A}} g_{\mu\nu}  A_{\beta} \nabla^\beta {\cal A}~, \\
    F_{\mu \nu} = & \nabla_\mu A_\nu - \nabla_\nu A_\mu~, \\
    {\cal B}_{\mu\nu}^\phi = & \nabla_\mu \nabla_\nu \phi - \frac{1}{{\cal A}} \nabla_{(\mu } {\cal A} \nabla_{\nu)} \phi + \frac{1}{2 {\cal A}} g_{\mu\nu} \nabla^\beta {\cal A} \nabla_\beta \phi~,\label{eq:Bmunuphi}
\end{align}
where the round brackets denote normalized symmetrization of indexes.
The conformal invariance is ensured by the overall factor ${\cal A}$, but it can be replaced by either $X$ or ${\cal Q}$. 
We can also define a conformal invariant parity breaking operator as
\begin{align}
    \tilde F^{\mu\nu}_C = \frac{\epsilon^{\mu\nu\alpha\beta}}{{\cal A}^2} F_{\alpha\beta}~.
\end{align}
However, in the following, we will assume parity conservation. 
Lastly, the conformal invariant curvature can be expressed as
\begin{align}\label{eq:DDD}
    {\cal D}_{\mu\nu\alpha\beta} = \chi_a R_{\mu\nu\alpha\beta} + g_{\beta [\mu} \nabla_{\nu]} \nabla_\alpha \chi_a - g_{\alpha [\mu} \nabla_{\nu]} \nabla_\beta \chi_a - \frac{3}{{\cal A}} \nabla_{[\alpha} \chi_a g_{\beta] [\mu} \nabla_{\nu]} \chi_a + \frac{\nabla_\lambda \chi_a \nabla^\lambda \chi_a}{2 \chi_a} g_{\beta [\mu} g_{\nu] \alpha}~\,,
\end{align}
where $\chi_a=\{X,{\cal A},{\cal Q}\}$. Note that we are free to consider ${\cal D}_{\mu\nu\alpha\beta}$ with $\chi_a={\cal A}$ only in \eqref{eq:DDD}. Other choices for $\chi_a$ can be recovered by adding or subtracting conformal invariant tensors involving second derivatives of ${\cal X}$ or ${\cal P}$. These can be obtained by replacing $\phi$ in $B_{\mu\nu}^\phi$ in \eqref{eq:Bmunuphi} for ${\cal X}$ or ${\cal P}$. Thus, from now on, we only consider ${\cal D}_{\mu\nu\alpha\beta}$ with $\chi_a={\cal A}$ without loss of generality. Lastly, it is important to note that to preserve the conformal invariance, matter must couple to the conformal invariant metric ${\cal G}_{\mu\nu}$ or a rescaled version of it, e.g. ${\cal X} {\cal G}_{\mu\nu}$.  In general, each matter sector could be coupled to a different conformal invariant metric.  

Armed with our building blocks, we write the leading order ( linear in the curvature and up to second order in ${\cal B}_{\mu\nu}^{\phi/A}$) conformal invariant action as
\begin{align}
    S_{\rm conf.} = \int \md^4x\, {\cal V} \Big[c_0 +  & c_1 {\cal D}_{\mu\nu\alpha\beta} {\cal G}^{\mu\alpha} {\cal G}^{\nu\beta} + 
    ( c_2 {\cal B}_{\mu\nu}^A + c_3 {\cal B}_{\mu\nu}^\phi ) {\cal C}_1^{\mu\nu}  +   c_4 F_{\mu\nu} C_2^{\mu\nu}  \nonumber \\& +\big( c_5 {\cal B}_{\mu\nu}^A {\cal B}^A_{\alpha\beta} + c_6 {\cal B}^A_{\mu\nu} {\cal B}^\phi_{\alpha\beta} + c_7 {\cal B}^\phi_{\mu\nu} {\cal B}^\phi_{\alpha\beta} \big) {\cal C}_3^{\mu\nu\alpha\beta} \nonumber \\&
     + c_8 F_{\mu\nu} F_{\alpha\beta}  C_4^{\mu\nu\alpha\beta} + (c_9 F_{\mu\nu} {\cal B}_{\alpha\beta}^A + c_{10} F_{\mu\nu} {\cal B}_{\alpha\beta}^\phi ) C_5^{\mu\nu\alpha\beta}  \big]~,
    \label{eq:Conformal_Vector_Scalar}
\end{align}
where $c_i=c_i(\phi,{\cal X},{\cal P})$, with $i=[1,10]$, are free functions and the tensors $C_j$ contain projections along all distinct directions, namely
\begin{align}
    C_1^{\mu\nu} = & d_1 {\cal G}^{\mu\nu} + d_2 {\cal N}^\mu_A {\cal N}^\nu_A + d_3 {\cal N}_A^{(\mu} {\cal N}^{\nu)}_\phi  + d_4 {\cal N}_\phi^\mu {\cal N}_\phi^\nu~, \\
    C_2^{\mu\nu} = & d_5  {\cal N}_\phi^{[\mu} {\cal N}^{\nu]}_A~,
\end{align}
where $d_i = d_i(\phi, {\cal X},{\cal P})$ are again free functions. Similarly, the other tensors $C_3^{\mu\nu\alpha\beta}$, $C_4^{\mu\nu\alpha\beta}$ and $C_5^{\mu\nu\alpha\beta}$ can be constructed out of $C_1^{\mu\nu}$ and $C_2^{\mu\nu}$ using that $C_3$ ($C_4$) is symmetric (antisymmetric) under exchange of $\mu \leftrightarrow \nu$ and $\alpha\leftrightarrow \beta$ and $C_5$ is symmetric under $\alpha \leftrightarrow \beta$ but antisymmetric under $\mu \leftrightarrow \nu$.

\subsection{Conformal Gauge Fixing}

After building general conformal invariant vector-scalar theories, let us study the different ways of fixing the conformal symmetry. Let us start from the conformal invariant theory given by
\begin{align}
    S_{\rm conf} =  \int \md^4x\, {\cal V} \big[ {\cal L}&( {\cal G}_{\mu\nu}(\phi, A_\alpha), \phi, A_\nu) \nonumber\\&
    +\lambda_X ( X + g^{\mu\nu} \nabla_\mu \phi \nabla_\nu \phi ) + \lambda_{\cal A} ( {\cal A} + g^{\mu\nu} A_\mu A_\nu ) +  \lambda_{\cal Q} ( {\cal Q} + g^{\mu\nu} \nabla_\mu \phi A_\nu )  \big]~,
\end{align}
where we add the Lagrange multiplier terms so as to treat $X$, ${\cal A}$ and ${\cal Q}$ as independent fields. Note that, for the moment, we do not use $X$, ${\cal A}$ and ${\cal Q}$ in the Lagrangian ${\cal L}$ so that, at the moment, they are just "dummy" variables. In addition, we assume that the Lagrange multiplier scale as $\lambda_a \rightarrow \Omega^2 \lambda_a $ under conformal transformations given by \eqref{eq:conf_transf}, with $a=\{1,2,3\}$. In this way, the conformal invariance is manifest. Note that by adding the Lagrange multipliers we do not change the equations of motion, as variation with respect to $\chi_a=\{X,{\cal A},{\cal Q}\}$ enforces $\lambda_a =0$. 

As the original action is conformal invariant, only the part proportional to the Lagrange multiplier is modified. Namely, after a conformal transformation given by \eqref{eq:conf_transf}, we find that the action reads
\begin{align}
    S_{\rm conf} =  \int \md^4x\,& {\cal V} \bigg\{ {\cal L}( {\cal G}_{\mu\nu}(\phi, A_\alpha), \phi, A_\nu)\nonumber\\& +\lambda_X \left( \Omega^{2} {X} + g^{\mu\nu} \nabla_\mu \phi \nabla_\nu \phi \right) + \lambda_{\cal A} \left( \Omega^{2} {{\cal A}}+ g^{\mu\nu} A_\mu A_\nu \right) +  \lambda_{\cal Q} \left( \Omega^{2} {{\cal Q}} + g^{\mu\nu} \nabla_\mu \phi A_\nu \right)  \bigg\}~.
\end{align}
Assuming that $\chi_a >0$, we can see that by fixing $\Omega^{2} = \chi_a^{-1}$ we obtain different gauge choices corresponding to $\chi_a =1$. Thus, the different constraints correspond to different frames. 

As an example, let us consider the mimetic constraint by choosing $\chi_a= X$. Further, we can redefine ${\cal A} \rightarrow {\cal A} X$ and ${\cal Q} \rightarrow {\cal Q} X$, as they are "dummy" variables, so that we obtain
\begin{align}
     S_{\rm conf} =  \int \md^4x\, {\cal V} \bigg\{ &{\cal L}( {\cal G}_{\mu\nu}(\phi, A_\alpha), \phi, A_\nu)\nonumber\\& +\lambda_X \left(1 + g^{\mu\nu} \nabla_\mu \phi \nabla_\nu \phi \right) + \lambda_{\cal A} \left( {\cal A} + g^{\mu\nu} A_\mu A_\nu \right) +  \lambda_{\cal Q} \left( {\cal Q} + g^{\mu\nu} \nabla_\mu \phi A_\nu \right)  \bigg\}~,
\end{align}
and we recover the mimetic constraint. The other two Lagrange multiplier terms can be dropped as they still do not modify the equations of motion. Similarly, we can obtain the TeVeS constraint for $\chi_a= {\cal A}$. Another popular choice in the literature is given by $f(\phi) A_\mu A^\mu + 1 =0$ \cite{Chaichian:2014qba,Benisty:2021cin}.
The different gauge choices are analyzed in more detail in section \ref{sec:Aether-Scalar-Gauge-Constraints} for the AeST theory around the FLRW background. 

\subsection{Relativistic MOND}
\label{subsec:Relativistic_MOND}

We end this section by showing that the conformal invariant models constructed above contain the AeST and TeVeS theories.

\textit{Aether-scalar-tensor theory:}
We have already studied in section \ref{sec:Preliminaries} how to obtain AeST theory as a conformal invariant model ( see Eq. \eqref{eq:newactionaether2} and the discussion below). For completeness reasons, we reformulate it in terms of the basic building blocks leading to
\begin{align}
    S_{\text{\AE},\rm conf. }= \int \md^4x\, \frac{\sqrt{-g} {\cal A}^2 }{16 \pi \tilde G} \bigg\{ &\left( {\cal D}_{\mu\nu\alpha\beta} - \frac{K_B}{2} F_{\mu\nu} F_{\alpha\beta} \right) {\cal G}^{\mu\alpha} {\cal G}^{\nu\beta} \nonumber\\& +2(2-K_B) F_{\mu\nu} {\cal N}_A^\mu {\cal N}_\phi^\nu + (2- K_B) \left({\cal X} - {\cal P}^2\right)   - {\cal F}( {\cal X}, {\cal P}) \bigg\}\,.
    \label{eq:Conformal_Invariant_Aether_Scalar}
\end{align}
Writing the action explicitly in terms of the Ricci scalar and derivatives of ${\cal A}$ yields
\begin{align}
    S_{\text{\AE},\rm conf. }= 
    \int \md^4x\, \frac{\sqrt{-g}  }{16 \pi \tilde G} \bigg\{  &{\cal A} R +  \frac{3}{2 {\cal A}} \nabla_\mu {\cal A} \nabla^\mu {\cal A}   - \frac{K_B}{2} F_{\mu\nu} F^{\mu\nu}   \nonumber \\
    & +2(2-K_B) F^{\mu\nu} A_\mu \nabla_\nu \phi+ (2- K_B) {\cal A}^2 \left({\cal X} - {\cal P}^2\right)   - {\cal A}^2 {\cal F}( {\cal X}, {\cal P}) \bigg\}~.
    \label{eq:Conformal_Invariant_Aether_Scalar}
\end{align}
If we fix the conformal gauge via introducing the constraint ${\cal A} =1$, it leads exactly to the AeST theory as in eq. \eqref{eq:aethereqA}. 

\textit{TeVeS:}
The action of the TeVeS theory \cite{Bekenstein:2004ne} is given by
\begin{align}
    S_{\rm TeVeS} = \frac{1}{2} \int \md^4x\, \sqrt{-g} \Big[  R - \frac{\kappa}{2} F_{\mu\nu} F^{\mu\nu} +  \lambda ( {\cal A} - 1 ) + \mu ( X - {\cal Q}^2) - V(\mu)  \Big] + S_M( \tilde g_{\mu\nu},\Psi_M)~,
\end{align}
where the matter sector is disformally coupled to
\begin{align}
    \tilde g_{\mu\nu} = e^{-2 \phi} g_{\mu\nu} -2 \sinh(2 \phi) A_\mu A_\nu~.
\end{align}
Similar to the previous case, we express the gravity part in a conformal invariant theory 
\begin{align}
    S_{\rm conf}^{\rm TeVeS} = \frac{1}{2}  \int \md^4x\,  \sqrt{-g}  \Big[ {\cal A} R +  \frac{3}{2 {\cal A}} \nabla_\mu {\cal A} \nabla^\mu {\cal A}  + {\cal A}^2 \mu ({\cal X} - {\cal P}^2 ) -{\cal A}^2 V(\mu) \Big] + S_M((\tilde g_{\rm conf.})_{\mu\nu},\Psi_M)~,
\end{align}
where the matter sector is coupled to
\begin{align}
    (\tilde g_{\rm conf})_{ \mu \nu } =& e^{-2 \phi } {\cal A} g_{\mu\nu} - 2 \sinh(2 \phi) A_\mu A_\nu~.
\end{align}
Fixing the conformal symmetry by imposing ${\cal A}=1$ we recover the TeVeS action. Of course, we can also choose any other gauge fixing constraint. 
Note, however, that in a cosmological background, the sign of $\dot \phi$ changes from matter to $\Lambda$ domination \cite{Skordis:2005xk,Skordis_2009}, and, therefore, it is not possible to use the mimetic constraint to fix the conformal symmetry during the whole cosmological evolution. Instead, fixing the mimetic constraint yields two branches in the solutions for the cosmological evolution. This is different to the AeST theory, where in a cosmological background $X>0$ remains time-like.

\section{Singular Disformal Transformation}
\label{sec:Singular_Disformal_transformation}

In this section, we demonstrate that the gauge-fixed conformal invariant action developed in Sec.~\eqref{sec:Conformal_Invariance} can also be derived from a singular disformal transformation. A generic disformal transformation depending on a scalar field $\phi$ and a vector field $A_\mu$ can be written as
\begin{align}
    \tilde g_{\mu\nu} = D_0 g_{\mu\nu} + D_1 \nabla_\mu \phi \nabla_\nu \phi +  D_2 A_\mu A_\nu + 2 D_3 \nabla_{(\mu} \phi A_{\nu)}~,
\end{align}
where $D_i = D_i(\phi,X,{\cal A},{\cal Q})$ with $i=[0,3]$.  The metric inverse is given by 
\begin{align}
    \tilde g^{\mu\nu} = \frac{1}{D_1}  g^{\mu\nu} + D_{1I}(x) \nabla^\mu \phi \nabla^\nu \phi + D_{2I}(x) A^\mu A^\nu + 2 D_{3I}(x) A^{(\mu} \nabla^{\nu)}\phi~,
\end{align}
where we defined
\begin{align}
    D_{1I} &= - \frac{ D_0 D_1 - D_1 D_2 {\cal A} + D_3^2 {\cal A}  }{D_0 T }\,\\
     D_{2I} &= - \frac{D_0 D_2 - D_1 D_2 X + D_3^2 X  }{D_0 T}\,,\\
     D_{3I} &= - \frac{D_0 D_3 + D_1 D_2 {\cal Q} - D_3^2 {\cal Q}}{D_0 T}~\,
\end{align}
and 
\begin{align}
    T = D_0^2 - (D_1 X + D_2 {\cal A} + 2 D_3 {\cal Q} ) D_0 + (X {\cal A} - {\cal Q}^2) (D_1 D_2 - D_3^2)~.
\end{align}
The disformal transformation can be inverted if the Jacobian is not singular. Following Ref.~\cite{Firouzjahi:2018xob} we can look at the eigenvalue equation of the determinant of the Jacobian
\begin{align}
\label{eq:Jacobian}
    \left(\frac{\partial \tilde g_{\mu\nu}}{\partial g_{\alpha\beta}} - \lambda^{(n)} \delta^\alpha_\mu \delta^\beta_\nu \right) \xi^{(n)}_{\alpha\beta} = 0~,
\end{align}
see appendix \ref{app:Eigenvalue_problem} for more details. Note that our disformal transformation is a generalization of the two-scalar field disformal transformation studied in Ref.~\cite{Firouzjahi:2018xob} by replacing one scalar field gradient with a vector field. Their detailed analysis is straightforwardly generalized to our case. However, analyzing the full disformal transformation is quite involved. Therefore, we will focus on two simple subcases separately.

\subsection{Pure Conformal Transformation}
In the case of pure conformal transformation $D_1= D_2 = D_3=0$ and $D_0 = D_0(\phi,X,{\cal A},{\cal Q})$ the conditions for a singular Jacobian are simply given by $\lambda^{(C)}= D_0$ with multiplicity 9 and
\begin{align}
 \lambda_\star =    D_0 - \left( D_{0,X} X + D_{0,{\cal A}} {\cal A} + D_{0,{\cal Q}} {\cal Q} \right)~, \qquad \xi_{\alpha\beta}^\star = g_{\alpha\beta}~.
\end{align}
The general solution is given by 
\begin{align}
    D_0 = {\cal A} f(\phi, {\cal X},{\cal P})~.
\end{align}
Note that this coincides with \eqref{eq:Metric_Conformal_Invariance} and the resulting theory is conformal invariant. The transformation can be straightforwardly generalized to disformal transformations of the type
\begin{align}
    \tilde g_{\mu\nu} = {\cal A} f(\phi,{\cal X},{\cal P}) g_{\mu\nu}+ f_1(\phi,{\cal X},{\cal P}) \nabla_\mu \phi \nabla_\nu \phi + f_2(\phi,{\cal X},{\cal P}) A_\mu A_\nu + 2 f_3(\phi,{\cal X},{\cal P}) \nabla_{(\mu} \phi A_{\nu)}~.
\end{align}
The disformal metric remains conformal invariant and is indeed the most general solution to the eigenvalue equation for the eigentensor $\xi_{\mu\nu}^\star=g_{\mu\nu}$.  As for the scalar case, the conformal symmetry can be linked to a non-invertible disformal transformation. Therefore, we could understand the relativistic MOND models as arising from a non-invertible disformal transformation similar to the mimetic matter models \cite{Chamseddine:2013kea}, which are conformal invariant. 

\subsection{Pure disformal Invariance}
For a pure scalar disformal transformation, it is known that pure conformal invariance is equivalent to pure disformal invariance \cite{Jirousek:2022rym,Domenech:2023ryc}. Let us  then consider for simplicity invariance under pure vector disformal transformation, which corresponds to the eigentensor 
\begin{align}
    \xi^\star_{\mu\nu} = A_\mu A_\nu~.
\end{align}
Substituting it into the eigenvalue equation leads to two conditions
\begin{align}
    & {\cal A} {\cal Q} D_{a \neq 2,X} + {\cal A}^2 D_{a \neq 2,{\cal A}} + {\cal Q}^2 D_{a\neq 2,{\cal Q}} =0~, \\
    & D_0 + {\cal A} {\cal Q} D_{2,X} + {\cal A}^2 D_{2,{\cal A}} + {\cal Q}^2 D_{2,{\cal Q}} = \lambda_\star~,
\end{align}
with the general solutions
\begin{align}
    D_2 = \frac{D_0(\phi,{\cal X}_{1,d},{\cal P}) }{{\cal A}} - d_2(\phi,{\cal X}_{1,d},{\cal P}), \qquad D_{a\neq 2} = D_{a\neq 2}(\phi,{\cal X}_{1,d},{\cal P})~,
\end{align}
where
\begin{align}
    {\cal X}_{1,d} = X - \frac{{\cal Q}^2}{{\cal A}} + \frac{{\cal Q}^2}{{\cal A}^2}~, \qquad {\cal P} = \frac{{\cal Q}}{{\cal A}}~,
\end{align}
are disformal invariant. The well-known (singular) disformal invariant metric 
\begin{align}
    \tilde g_{\mu\nu} = g_{\mu\nu} + \left( \frac{1}{{\cal A}} -1 \right) A_{\mu} A_\nu
\end{align}
corresponds to the simplest subcase, which does not have $\phi$ dependence.

The previous results can also be straightforwardly generalized to scalar disformal invariance by considering the eigentensor $\xi^\star_{\mu\nu} = \nabla_\mu \phi \nabla_\nu \phi$, leading instead to 
\begin{align}
    \tilde g_{\mu\nu} = d_0 g_{\mu\nu} + \left( \frac{d_0}{X} - d_1 \right) \nabla_\mu \phi \nabla_\nu \phi  + d_2 A_\mu A_\nu + 2 d_3 \nabla_{(\mu} \phi A_{\nu)}~,
\end{align}
where the free functions now depend on $d_i = d_i(\phi, {\cal X}_{2,d},{\cal P}_2)$ with ${\cal X}_{2,d} = {\cal A} - {\cal Q}^2/X +{\cal Q}^2/ X^2$ and ${\cal P}_2= {\cal Q}/X$. 

Similarly to mimetic gravity models, the disformal invariant model is equivalent to the conformal one, which is shown explicitly in the appendix \ref{app:Disformal_Invarianz}. We expect that following \cite{Jirousek:2022rym}, any singular point in the Jacobian can be deformed to the conformal symmetry, which is, however, beyond the scope of this paper.

\section{Aether-scalar-tensor theory and the gauge constraints}
\label{sec:Aether-Scalar-Gauge-Constraints}

To get a first-hand understanding of the symmetries and gauge fixing constraints it is interesting to revisit the AeST theory in a FLRW background. Instead of using the original formulation, however, we will discuss it in the conformal invariant formulation and show the impact of different gauge choices. In the first part, we discuss the FLRW background in the Hamiltonian framework.

\subsection{Hamiltonian in FLRW background}
Using the FLRW metric at the background level, we can make the ansatz
\begin{align}
    \md s^2 = - N^2 \md t^2 + a^2 \md x^j \md x^i \delta_{ij}, \qquad \phi = \phi(t), \qquad A_\mu = A_0(t) \delta_\mu^0~.
\end{align}
With our anstaz, Eq.~\eqref{eq:Conformal_Invariant_Aether_Scalar} the action simplifies to
\begin{align}
    S_{\text{\AE,\rm conf}}=& \int \md^4x\, \frac{a^3 N}{16 \pi  G} \big[ 6 \frac{A_0^2}{N^2} \left( \frac{\md }{N a \md t}\left( \frac{\dot a}{N} \right) + \frac{\dot a^2 }{a^2 N^2} \right) - \frac{3}{ 2 A_0^2 } \left(\frac{\md}{\md t} \left( \frac{A_0}{N}\right)^2\right)^2  - \frac{A_0^4}{N^4} {\cal F}( 0, \dot \phi/A_0) \big] \nonumber \\
    =& \int \md^4x\, \frac{a^3 A_0^2}{16 \pi G N^3} \left[ - 6 \left( \frac{\dot a}{a} + \frac{\dot A_0}{A_0} - \frac{\dot N}{N}\right)^2 + {\cal K}_2 (- {\cal Q}_0 A_0 + \dot \phi)^2 - \Lambda A_0^2 \right]~.
\end{align}
where in the second step we have assumed the standard cosmological expansion for ${\cal F}$ \cite{Skordis:2020eui}
\begin{align}\label{eq:Fdef}
    {\cal F}(0,{\cal P}) \simeq  \Lambda -  {\cal K}_2 ({\cal P}- {\cal Q}_0)^2~.
\end{align}
By introducing a rescaled scale factor $\tilde a = a A_0/ N$, the lapse function drops out, and its role is replaced by $A_0$. This is similar to what happens for mimetic gravity, where the role of the lapse function is replaced by $\dot \phi$ (see, for instance, \cite{Domenech:2023ryc}).

As a next step, we study the Hamiltonian to get a better understanding of the symmetries. The canonical momenta are
\begin{align}
    p_a = - \frac{3 a^2 A_0^2 }{4 \pi  G N^3 } \left( \frac{\dot a}{a} +  \frac{\dot A_0}{A_0} - \frac{\dot N}{N}\right) \quad,& \quad
     p_A =   - \frac{3 a^3 A_0 }{4 \pi  G N^3 }  \left( \frac{\dot a}{a} +  \frac{\dot A_0}{A_0} - \frac{\dot N}{N}\right)~, \\
     p_N=   \frac{3 a^3 A_0^2 }{4 \pi  G N^4 }  \left( \frac{\dot a}{a} +  \frac{\dot A_0}{A_0} - \frac{\dot N}{N}\right)\quad,&\quad 
    p_\phi = \frac{a^3 A_0^2 {\cal K}_2 }{8 \pi  G N^3} \left( - {\cal Q}_0 A_0 + \dot \phi \right)~.
\end{align}
There are two primary constraints given by
\begin{align}
    C_1 = a p_a  + N p_N  \approx 0~, \\
    C_2 = - a p_a + A_0 p_A \approx 0~,
\end{align}
where, as usual, $\approx$ denotes equality on the constrained hypersurface. 
Performing the full Hamiltonian analysis, we obtain
\begin{align}
    H_T = \int \md^3x\, \big[ u_1 {\cal H} + u_2 C_1 + u_3 C_2 \big]~,
\end{align}
where the Hamiltonian density is given by
\begin{align}
    {\cal H} = {\cal Q}_0 A_0 p_\phi + \frac{4 \pi  G N^3 p_\phi^2}{a^3 {\cal K}_2 A_0^2} - \frac{2 \pi  G N^3 p_a^2 }{3 a A_0^2 } + \frac{\Lambda a^3 A_0^4 }{16 \pi  G N^3}~.
\end{align}

We find that all the constraints are first-class, resulting in one dynamical degree of freedom. The first-class Hamiltonian constraint is connected to the time reparametrization invariance, and $C_1$ and $C_2$ to the freedom in fixing the lapse function and the conformal symmetry. 
It is straightforward to check that the gauge invariant quantities introduced in the previous sections, namely  
\begin{align}
     \sqrt{{\cal X}}={\cal P}= \frac{\dot \phi}{A_0} = {\cal Q}_0  + \frac{8 \pi  G N^3 p_\phi}{{\cal K}_2 a^3 A_0^3}\,,
\end{align}
are indeed conformal invariant as they commute with $C_1$ and $C_2$. The standard gauge fixing conditions for the lapse function and the conformal freedom correspond to $A_0=1$ and $N=1$. In this case, we are back to the standard AeST theory. 

Let us now turn to the two different gauge fixing constraints $X=1$ or ${\cal Q}=1$. The mimetic constraint, $X=1$, corresponds to adding
\begin{align}
    C_{\rm mim} \equiv \frac{\dot \phi}{N} -1  = \frac{{\cal Q}_0 A_0}{N}  + \frac{8 \pi  G N^2 p_\phi}{{\cal K}_2 a^3 A_0^2} -1 \approx 0
\end{align}
to the Hamiltonian to fix the first class constraint $C_1$. 
While the constraint still commutes with $C_1 + C_2$, it does not directly commute with the Hamiltonian constraint, but we can redefine the constraint via
\begin{align}
    \tilde {\cal H} = {\cal H} - \frac{32 \pi^2 G^2  N^6 p_a p_\phi }{a^2 A_0^2 ( a^3 A_0^3 {\cal K}_2 {\cal Q}_0 + 8 \pi  G N^3 p_\phi ) } C_2~,
\end{align}
so that we keep two first class $C_1+ C_2$, $\tilde {\cal H}$ and two second class constraint $C_1$, $C_{\rm mim}$.

The other gauge choice ${\cal Q}=1$ corresponds in the FLRW background to 
\begin{align}
    C_{{\cal Q}} \equiv \frac{\dot \phi A_0}{N^2} - 1 = \frac{{\cal Q}_0 A_0^2}{N^2} + \frac{8 \pi  G N p_\phi}{{\cal K}_2 a^3 A_0} -1 \approx 0~.
\end{align}
As before, the constraint commutes with $C_1 + C_2$. Similarly, we can again redefine the Hamiltonian constraint to keep its first-class nature
\begin{align}
     \tilde {\cal H} = {\cal H} - \frac{16 \pi^2  G^2  N^6 p_a p_\phi }{a^2 A_0^2 ( a^3 A_0^3 {\cal K}_2 {\cal Q}_0 + 8 \pi  G N^3 p_\phi ) } C_2~,
\end{align}
so that we keep two first-class constraints after fixing the conformal gauge symmetry. 

In summary, adding the constraints $X=1$ or ${\cal Q}=1$ to the Hamiltonian is indeed nothing else but a gauge fixing constraint of the conformal symmetry.

\subsection{Equations of motion}
As a next step, we consider the background equations of motion. We consider the gauge fixing condition ${\cal Q}=1$, which we add directly via a Lagrange multiplier term to the action. As the derivative coupling between the scalar field and the vector field vanishes in the FLRW background, after introducing the gauge fixing constraint ${\cal Q}=1$ the original action does not depend anymore on directly on $\dot \phi$. We obtain
\begin{align}
     S= S_{\text{\AE,\rm conf}} + \int \md^4x\, \sqrt{-g}& {\cal A}^2 \lambda ({\cal Q} -1)
     = \int \md^4x\, \frac{a^3 N}{16 \pi G} \bigg\{ 6 \frac{A_0^2}{N^2} \left( \frac{\md }{N a \md t}\left( \frac{\dot a}{N} \right) + \frac{\dot a^2 }{a^2 N^2} \right) \nonumber\\&- \frac{3}{ 2 A_0^2 } \left(\frac{\md}{\md t} \left( \frac{A_0}{N}\right)^2\right)^2  - \frac{A_0^4}{N^4} {\cal F}( 0, 1/{\cal A})  \} + \frac{A_0^4 a^3}{N^3}  \lambda \left( \frac{A_0 \dot \phi}{N^2} - 1 \right)~.
\end{align}
Recall that ${\cal F}$ is given in Eq.~\eqref{eq:Fdef}. Further, it is useful to introduce the rescaled scale factor $\tilde a = a A_0/ N$, which leads to
\begin{align}
    S = \int \md^4x\, \frac{\tilde a^3}{16 \pi  G A_0^3} \bigg\{ A_0^4 ({\cal K}_2 {\cal Q}_0^2 - \Lambda) + {\cal K}_2 N^4 - 2 A_0^2  (3 \tilde H^2 + {\cal K}_2 {\cal Q}_0 N^2 ) \bigg\} + \frac{A_0 \tilde a^3 \lambda}{N^2} ( A_0 \dot \phi - N^2)~,
\end{align}
where $\tilde H = \dot{\tilde a}/\tilde a$.
The equation of motion for $\dot \phi$ and the lapse become very simple, namely
\begin{align}
    \tilde a^3 \lambda A_0^2 = C_0\quad,\quad
     \lambda =  \frac{1 }{16 \pi  G A_0^3 \dot \phi} \frac{ \partial {\cal F}(0,1/{\cal A})}{\partial (1/{\cal A}) } =\frac{{\cal K}_2 (1- {\cal Q}_0 A_0^2 )}{8 \pi \tilde G A_0^4}~, \label{eq:lambda_solution_background}
\end{align}
where we fixed the cosmic frame $N=1$ and $C_0$ is an integration constant. There is the trivial solution $C_0=0$, $A_0 = 1/\sqrt{{\cal Q}_0}$ and $\lambda =0$, which leads to the standard de-Sitter solution as there is only a cosmological constant. On the other hand, for $C_0 \neq 0$ we get
\begin{align}
    A_0 =  \frac{\sqrt{{\cal K}_2 \tilde a^3} }{\sqrt{8 \pi  G C_0 + {\cal K}_2 {\cal Q}_0 \tilde a^3}}~\quad{\rm and}\quad
    \lambda =  \frac{8 \pi  G C_0^2 + C_0 {\cal K}_2 {\cal Q}_0 \tilde a^3}{\tilde a^6 {\cal K}_2}~.
\end{align}
Lastly, the two Friedmann equations are given by
\begin{align}
     \frac{3 \tilde H^2}{8 \pi  G} = &   \frac{C_0}{\tilde a^3 } + \frac{\Lambda }{  16 \pi  G {\cal Q}_0} + \frac{4 \pi G C_0^2 }{{\cal K}_2 {\cal Q}_0 \tilde a^6} - \frac{3 C_0 \tilde H^2 }{{\cal K}_2 {\cal Q}_0 \tilde a^3 }~, \\
    \frac{3 \tilde H^2 + 2 \dot{\tilde H}}{8 \pi G} = & \frac{\Lambda}{16 \pi  G {\cal Q}_0 } - \frac{4 \pi G C_0^2  }{{\cal K}_2 {\cal Q}_0 \tilde a^6 } - \frac{2 C_0 \dot{\tilde H} }{{\cal K}_2 {\cal Q}_0 \tilde a^3}~.
\end{align}
Assuming $C_0 \ll 1$, we can perturbatively solve the Friedmann equations in powers of $C_0$, so that we recover the standard equations of motion for a $\Lambda$CDM background at leading order. We identify the leading order correction from the integration constant as the CDM contribution, and its energy density is given by 
\begin{align}
    \rho_{CDM}\equiv \frac{C_0}{\tilde a^3} = \frac{\lambda}{{\cal Q}_0} + {\cal O}(C_0^2)~.
\end{align}
Obviously, we recover the same solution as for the standard constraint ${\cal A}=1$, but the freedom in choosing the gauge fixing constraint can simplify the calculations and provide an alternative viewpoint. Note that in contrast to standard mimetic matter, in the present case, the mimetic constraint by itself does not lead to a dust component. The reason for that is that in the case of ${\cal F}=0$, the gravity part is conformal invariant, and variation with respect to the lapse yields vanishing $\lambda$ (see Eq. \eqref{eq:lambda_solution_background}).

\section{Conclusion}
\label{sec:Conclusion}

In this work, we have explored the connection between relativistic MOND theories and mimetic gravity. We demonstrated that any relativistic MOND model featuring a unit time-like vector field, such as TeVeS or AeST theories, can be formulated within a conformal-invariant framework. The conformal invariant formulation reveals that the constraints imposed in TeVeS and mimetic gravity emerge as different gauge-fixing choices, highlighting the conformal symmetry underlying both theories. The different gauge-fixing constraints correspond to a change of frame and can be interchanged via a conformal transformation, provided that the derivative of the scalar field remains time-like. 

In particular, we have shown that the TeVeS constraint given by $A_\mu A^\mu=-1$ can be interchanged by $\nabla_\mu\phi\nabla^\mu\phi=-1$, or $A^\mu\nabla_\mu\phi=-1$. The resulting action for $A_\mu$ then falls into a simple subclass of generalized Proca theories \cite{Heisenberg:2014rta} with an additional non-minimal coupling to gravity. While these are equivalent formulations of the theory, we believe that a different choice of constraints may lead to interesting interpretations of the MOND-like behavior in the weak field regime. It is also possible that in some cases, choosing one frame or another may result in the simplification of the equations of motion. As a concrete example, we examined the Hamiltonian structure of the AeST in an FLRW background and demonstrated how different constraints correspond to different gauge-fixing choices of the conformal symmetry.

We also have demonstrated how relativistic MOND theories can be linked to non-invertible disformal transformations. For instance, we have shown that these models can alternatively be reformulated in terms of disformal symmetry, similar to scalar-tensor theories \cite{Jirousek:2018ago, Domenech:2023ryc}, as both symmetries are equivalent and can be deformed into each other. 

By constructing the fundamental building blocks for a conformal-invariant vector-scalar-tensor theory, with TeVeS and AeST theory as specific cases, we established a new framework for developing relativistic MOND theories. It should be noted that not all conformal-invariant vector-scalar-tensor theories are suitable relativistic MOND theories. We leave as future work the classification of conformal-invariant vector-scalar-tensor theory that recovers MOND in the weak field regime. We also expect that by working with different conformal gauge-fixing choices, we will gain a new interpretation of the origin of the MOND-like behavior of these theories. For instance, although speculative, since choosing the mimetic constraint $X=1$ leads to a generalized Proca theory, it is plausible that the MOND-like behavior is linked to screening mechanisms like the Vainshtein mechanism. Furthermore, it would be interesting to investigate the linear perturbations around the FLRW background, particularly given that mimetic gravity models suffer from instability issues \cite{Langlois:2018jdg,Takahashi:2017pje,Zheng:2017qfs,Firouzjahi:2017txv,Ijjas:2016pad,Hirano:2017zox,Gorji:2017cai,Ganz:2018mqi,Ganz:2019vre}. Studying how the presence of a vector field can stabilize these perturbations may provide further insights into these models. 
A well-known issue in mimetic gravity is the formation of caustics \cite{Babichev:2017lrx}. Similarly, in TeVeS, it has been shown that the vector field can develop caustics due to the TeVeS constraint \cite{Contaldi:2008iw}. This raises the question of whether caustics are a generic problem in conformally invariant models and whether they can be linked to the TeVeS or mimetic constraints. Furthermore, it is worth exploring whether the choice of gauge-fixing constraint allows for shifting caustics between the scalar and vector fields.
It would also be interesting to study whether the vector field could lead to a sizeable production of gravitational waves in the early universe, as,e.g., the vector dark matter model recently studied in Ref.~\cite{Marriott-Best:2025sez}.

We end by noting that there could be subtelties in mapping MOND and mimetic theories when $\dot\phi$ crosses zero, which, for instance, is the case in TeVeS. If so, we expect that the solutions of the theory in the mimetic constraint formulation will split into two different branches. Furthermore, in the weak-field regime of TeVeS, which is characterised by $\dot \phi \ll 1$, the scalar field may become spacelike, and, therefore, one must carefully choose the sign of the mimetic constraint. In contrast, in AeST theory, the scalar field enjoys shift symmetry, and $\dot \phi$ remains always time-like. We leave further investigations for future work.

\begin{acknowledgments}
We would like to thank M.~A.~Gorji, M.~Sasaki, G.~Tasinato and T.~Zlosnik for their helpful comments and discussions. This research is supported by the DFG under the Emmy-Noether program, project number 496592360, and by the JSPS KAKENHI grant No. JP24K00624. Calculations of the disformal and conformal invariant tensors were cross-checked with the Mathematica package xAct (www.xact.es).
\end{acknowledgments}

\appendix

\section{Disformal Building Blocks}
\label{app:Disformal_Invarianz}

Following the construction of the conformal building blocks in section \ref{sec:Conformal_Invariance} we can also construct models which are invariant under disformal transformation. For simplicity, we consider just one case, i.e. invariance under pure vector disformal transformations
\begin{align}
    \tilde g_{\mu\nu} = g_{\mu\nu} + D(x) A_\mu A_\nu~.
\end{align}
As derived in section \ref{sec:Singular_Disformal_transformation} the most general ansatz for a metric depending on $\phi$, $\chi_a$, which is invariant under pure-vector disformal transformations, is given by
\begin{align}
    ({\cal G}_d)_{\mu\nu} = D_0 g_{\mu\nu} + D_1 \nabla_\mu \phi \nabla_\nu \phi + \left( \frac{D_0}{{\cal A}} - d_2 \right) A_{\mu} A_{\nu} + 2 D_3 A_{(\mu} \nabla_{\nu)}\phi~,
\end{align}
where each of the free functions depend on $D_a = D_a(\phi,{\cal X}_{1,d},{\cal P})$. However, similarly to the conformal theories \ref{sec:Conformal_Invariance} we can use the most simplistic ansatz $D_0=1=d_2$, $D_1=D_3=0$ to construct all the disformal building blocks without loss of generality. These are given by 
\begin{align}
     {\cal V}_d = & \sqrt{- {\cal G}_d } = \sqrt{-g {\cal A}}~, \quad ({\cal N}_d)_A^\mu = \frac{A^\mu}{{\cal A}}~, \quad ({\cal N}_d)_\phi^\mu = {\cal G}_d^{\alpha \beta} \nabla_\mu \phi\,,
\end{align}
\begin{align}
{\cal P} = \frac{{\cal Q}}{{\cal A}}~, \quad {\cal X}_{1,d} = X - \frac{{\cal Q}^2}{{\cal A}} + \frac{{\cal Q}^2}{{\cal A}^2}~,
\end{align}
\begin{align}
    ({\cal B}_d)_{\mu\nu}^A = & \frac{1}{{\cal A}} \left( \nabla_{(\mu} A_{\nu)} - \frac{1}{{\cal A}} \nabla_{(\mu} {\cal A} A_{\nu)} - \frac{1}{2 {\cal A}^2} A^\alpha \nabla_{\alpha} {\cal A} A_\mu A_\nu - \frac{1}{{\cal A}} A_{(\mu} F_{\nu)\alpha} A^\alpha  \right) + A_{(\mu} F_{\nu)\alpha} ({\cal N}_d)^\alpha~, \\
    F_{\mu\nu} =& \nabla_\mu A_\nu - \nabla_\nu A_\mu~, \\
     ({\cal B}_d)_{\mu\nu}^\phi = &  \nabla_\mu \nabla_\nu \phi  - \frac{(1 - {\cal A})^2 {\cal Q} }{ {\cal A}^3} A^\alpha A_{(\nu} \nabla_\alpha A_{\mu)} + \frac{ (1- {\cal A})  {\cal Q}}{2 {\cal A}^4} A^\alpha \nabla_\alpha {\cal A} A_\mu A_\nu+ \frac{1- {\cal A} }{{\cal A}} \nabla^\alpha \phi F_{\alpha (\mu} A_{\nu)} \nonumber \\
   &
     - \frac{1}{2 {\cal A}^2} A_\mu A_\nu \nabla_\alpha \phi \nabla^\alpha {\cal A} + \frac{({\cal A} -1) {\cal Q} }{ {\cal A}^2} \nabla_{(\mu} A_{\nu)}  + \frac{(1 + 2 {\cal A} - {\cal A}^2) {\cal Q}}{2 {\cal A}^3} A_{(\mu } \nabla_{\nu)} {\cal A}~\,.
\end{align}
Note that we can also write $({\cal B}_d)_{\mu\nu}^\phi =   \bar\nabla_\mu \bar\nabla_\nu \phi$, where $\bar \nabla_\mu $ is the covariant derivative with respect to ${\cal G}_{\mu\nu}$.
The first term is explicitly constructed such that ${\cal B}_{\mu\nu}\vert_{{\cal A}=1}= \nabla_{(\mu} A_{\nu)}$. Further, we have can consider derivatives of the conformal invariant ratios, namely $\nabla_{\mu} {\cal X}_{1,d}$ and $\nabla_\mu {\cal P}$.

Lastly, we can construct the disformal invariant curvature tensor. 
As before, we can construct as $R_{\mu\nu\alpha\beta}[{\cal G}_d] = ({\cal D}_d)_{\mu\nu\alpha\beta } $, which is the Riemann tensor induced by the disformal invariant metric ${\cal G}_{\mu\nu}$.
As the expression is quite involved we will only write down the Ricci-scalar explicitly which is given by
\begin{align}
    {\cal D}_d\equiv & ({\cal D}_d)_{\mu\nu\alpha\beta} {\cal G}^{\mu\alpha}{\cal G}^{\nu\beta}=   R - \frac{\Box {\cal A}}{{\cal A}^2} + \frac{(1+  {\cal A})^2}{8 {\cal A}^3} \nabla_\mu {\cal A} \nabla^\mu {\cal A} + \frac{1-{\cal A}}{{\cal A}^2} \left(2 A^\mu \nabla_\mu \nabla_\nu A^\nu + (\nabla_\mu A^\mu)^2 \right) \nonumber \\
    &+ \frac{{\cal A}-3}{{\cal A}^3} A^\mu \nabla_\mu {\cal A} \nabla_\beta A^\beta  + \frac{(1-{\cal A})^2}{2 {\cal A}^3} A^\alpha A^\beta \nabla_\alpha A^\mu \nabla_\beta A_\mu - \frac{3+ 2 {\cal A} - {\cal A}^2 }{2 {\cal A}^3} A^\alpha \nabla_\alpha A^\beta \nabla_\beta {\cal A} \nonumber \\
    &+ \frac{5 }{2 {\cal A}^4} A^\alpha A^\beta \nabla_\alpha {\cal A} \nabla_\beta {\cal A} - \frac{1}{{\cal A}^3} A^\alpha A^\beta \nabla_\alpha \nabla_\beta {\cal A}  + \frac{2 ({\cal A}-1)}{{\cal A}^2} A^\alpha \nabla_\beta \nabla^\beta A_\alpha \nonumber \\
    &+ \frac{1- {\cal A}^2}{2 {\cal A}^2} \nabla_\alpha A^\beta \nabla_\beta A^\alpha + \frac{ {\cal A}^2 -3 + 2 {\cal A} }{2 {\cal A}^2} \nabla_\beta A^\alpha \nabla^\beta A_\alpha~.
\end{align}
By construction for ${\cal A}=1$ we obtain that all the disformal invariant operators coincide with the conformal ones, i.e.
\begin{align}
   ({\cal G}_d)_{\mu\nu} \big \vert_{{\cal A}=1} = {\cal G}_{\mu\nu} \big \vert_{{\cal A}=1}  , \qquad ({\cal B}_d)_{\mu\nu}^{\phi/A} \big \vert_{{\cal A}=1} =  {\cal B}_{\mu\nu}^{\phi/A} \big \vert_{{\cal A}=1}, \qquad ({\cal D}_d)_{\mu\nu\alpha\beta}\big \vert_{{\cal A}=1} = {\cal D}_{\mu\nu\alpha\beta}\big \vert_{{\cal A}=1}~,
\end{align}
etcetera.
Therefore, we can see that these theories conincide after choosing the conformal/ disformal gauge.
\begin{align}
   & S_{\rm conf.} + \int \md^4x\, \sqrt{-g}\, \lambda ({\cal A} -1 ) = \int \md^4x\,  \left( {\cal V} {\cal L}( {\cal G}_{\mu\nu}, {\cal D}_{\mu\nu\alpha,\beta}, ...) + \sqrt{-g} \lambda ({\cal A} -1 ) \right) 
    \nonumber \\
    & = \int \md^4x\,  \left( {\cal V}_d  {\cal L}( ({\cal G}_d)_{,\mu\nu}, ({\cal D}_d)_{\mu\nu\alpha\beta}, ...) +  \sqrt{-g} \lambda ({\cal A} -1 ) \right) = S_{\rm disf.} + \int \md^4x\, \sqrt{-g} \lambda ({\cal A} +1 )~.
\end{align}
The same approach can be done for pure scalar disformal transformation generalizing \cite{Domenech:2023ryc} which can also be deformed to conformal invariance. Therefore, we can see that disformal and conformal ivnariance are equivalent to each other, in the sense that after gauge fixing they lead to the same equations of motion. 

\if{}
\subsection{Scalar Disformal Transformation}
\label{app_subsec:Scalar_Disformal_Transformation}

Pure scalar disformal transformation has been discussed in details in \cite{Domenech:2023ryc}. It is possible to express all the curvature and derivatives of the scalar field in terms of the disformal metric
\begin{align}
    \tilde g_{\mu\nu} = g_{\mu\nu} + \left( \frac{1}{X} -1 \right) \phi_\mu \phi_\nu
\end{align}
Obviously, for the condition $X=1$ we end up with the original metric. Adding the vector field we can again define the two invariant contractions
\begin{align}
    {\cal X}_{2,d} = {\cal A} - \frac{{\cal Q}^2}{X} + \frac{{\cal Q}^2}{X^2}, \qquad {\cal P} = \frac{{\cal Q}^2}{X^2}
\end{align}
which follows directly from the vector case by replacing $X \leftrightarrow {\cal A}$.

The missing terms are the derivatives of the vector field which can be constructed in a similar fashion and is given by
\begin{align}
    ({\cal B}_{\phi})^A_{\mu\nu} = \tilde  \nabla_{(\mu} A_{\nu)} = ..., \\
    F_{\mu\nu} = \nabla_\mu A_\nu - \nabla_\nu A_\mu
\end{align}
To show the equivalence to the conformal invariant model we can note that we can replace the derivatives of ${\cal A}$ in the definition of the conformal invariant derivative of the vector field ${\cal B}_{\mu\nu}$ by using the vectors ${\cal B}_{1c,\mu}$, ${\cal B}_{1c,\nu}$, i.e.
\begin{align}
    {\cal B}_{\mu\nu}^A - \frac{1}{X_{1c}} {\cal B}_{1c,(\mu} A_{\nu)} + \frac{1}{2 X_{1c}} g_{\mu\nu} {\cal B}_{1c,\beta} {\cal N}^\beta_A = \nabla_{(\mu} A_{\nu)} - \frac{1}{X} \nabla_{(\mu} X A_{\nu)} + \frac{1}{2 X} g_{\mu\nu} A^\beta \nabla_\beta X
\end{align}
A similar procedure can be done for the derivative of the scalar field. For the Riemann tensor we need also $({\cal B}_{X_{a,c}})_{\mu\nu}$ but in combination we can replace any ${\cal A}$ by $X$ in the definition of ${\cal D}_{\mu\nu\alpha\beta} $.

Therefore, it is straightforward to see if we start from a general action in terms of the conformal invariant operators
\begin{align}
    S_{\rm conf} = \int \md^4x\, L( {\cal G}_{\mu\nu}, A_\mu, \phi, {\cal X}, {\cal P}, {\cal B}_{\mu\nu}^A, {\cal B}_{\mu\nu}^\phi, ... ) 
\end{align}
we can always replace any of the ${\cal A}$ or derivatives of it by $X$ and its derivatives and therefore, after gauge fixing $X =1$ we recover the same action but now with the mimetic constraint as if we start from the scalar disformal one just with reshuffled coefficients. 
\fi

\section{Eigenvalue problem}
\label{app:Eigenvalue_problem}
The eigenvalue equation for the disformal transformation eq. \eqref{eq:Jacobian} can be expressed as
\begin{align}
    ( D_0 -\lambda^{(n)}) \xi^{(n)}_{\mu\nu} + {\cal M}^{\alpha\beta}_{\mu\nu} \xi_{\alpha\beta}^{(n)} =0~,
\end{align}
where
\begin{align}
    {\cal M}_{\mu\nu}^{\alpha\beta} = & \left( D_{0,X} g_{\mu\nu} + D_{1,X} \nabla_\mu \phi \nabla_\nu \phi + D_{2,X} A_\mu A_\nu + 2 D_{3,X} \nabla_{(\mu}\phi A_{\nu)}  \right) \nabla^\alpha\phi \nabla^\beta \phi \nonumber \\
    & +  \left(D_{0,{\cal A}} g_{\mu\nu} + D_{1,{\cal A}} \nabla_\mu \phi \nabla_\nu \phi + D_{2,{\cal A}} A_\mu A_\nu + 2 D_{3,{\cal A}} \nabla_{(\mu}\phi A_{\nu)}  \right) A^\alpha A^\beta \nonumber \\
    & + \left( D_{0,{\cal Q}} g_{\mu\nu} + D_{1,{\cal Q}} \nabla_\mu \phi \nabla_\nu \phi + D_{2,{\cal Q}} A_\mu A_\nu + 2 D_{3,{\cal Q}} \nabla_{(\mu}\phi A_{\nu)}  \right) \nabla^{(\alpha} \phi A^{\beta)}~.
\end{align}
In order to solve the eigenvalue equation we can decompose the metric as
\begin{align}
    g_{\mu\nu} =  h_{\mu\nu} - \frac{\nabla_\mu \phi \nabla_\nu \phi}{X}~,
\end{align}
where we can further split
\begin{align}
    h_{\mu\nu} = \hat h_{\mu\nu} + \frac{h^{\alpha}_\mu A_\alpha h^{\beta}_\nu A_\beta}{A_\sigma A_\lambda h^{\sigma \lambda}}~.
\end{align}
Similarly, we can split
\begin{align}
    \xi^\star_{\mu\nu} = A \hat h_{\mu\nu} + B h^{\alpha}_\mu h^\beta_\nu A_\alpha A_\beta + C h^{\alpha}_{(\mu} A_\alpha \nabla_{\nu)} \phi + D \nabla_\mu \phi \nabla_\nu \phi~.
\end{align}
It is straightforward to see that there is one trivial solution of the eigenvalue eq. \eqref{eq:Jacobian}
\begin{align}
    \lambda^{(C)} = D_0~, \qquad \xi_{\mu\nu}^{(C)} = \hat h_{\mu\nu}~.
\end{align}
The solution for the eigentensor is degenerate with multiplicity of nine \cite{Firouzjahi:2018xob}.  The remaining solution is quite involved and, therefore, we focus on two simple cases: I) $D_1=D_2=D_3=0$ and II) $\xi^\star_{\mu\nu} = A_\mu A_\nu$ or $\xi^\star_{\mu\nu} = \nabla_\mu \phi \nabla_\nu \phi$ as discussed in section \ref{sec:Singular_Disformal_transformation}.

\bibliography{bibliography.bib} 

\end{document}